\documentclass[pra,twocolumn,superscriptaddress,10pt]{revtex4-1}
\usepackage{graphicx}
\usepackage{dcolumn}
\usepackage{color}
\usepackage{times}
\usepackage{bm}
\usepackage{amssymb}
\usepackage{amsmath}
\usepackage{epsfig}
\usepackage{epstopdf}
\usepackage{dsfont}
\usepackage{subfigure}
\usepackage{float}
\usepackage[colorlinks=true,citecolor=blue, linkcolor=blue, urlcolor=blue]{hyperref}
\begin{document}
\renewcommand{\thefootnote}{\fnsymbol {footnote}}
	
	\title{{Einstein-Podolsky-Rosen Steering Criterion and Monogamy Relation via Correlation Matrices in Tripartite Systems}}
	
	\author{Li-Juan Li}
	\affiliation{School of Physics and Optoelectronic Engineering, Anhui University, Hefei 230601, China}

 \author{Xiao Gang Fan}
	\affiliation{School of Physics and Optoelectronic Engineering, Anhui University, Hefei 230601, China}
   \author{Xue-Ke Song}
\affiliation{School of Physics and Optoelectronic Engineering, Anhui University, Hefei 230601, China}

	%\affiliation{CAS Key Laboratory of Quantum Information,University of Science and Technology of China, Hefei 230026, China}

	\author{Liu Ye}
	%\email{yeliu@ahu.edu.cn}
	\affiliation{School of Physics and Optoelectronic Engineering, Anhui University, Hefei 230601, China}
%\affiliation{Center for Quantum Information, IIIS, Tsinghua University, Beijing 100084, People's Republic of China}
	\author{Dong Wang}
	\email{dwang@ahu.edu.cn}
    \affiliation{School of Physics and Optoelectronic Engineering, Anhui University, Hefei 230601, China}
    %\affiliation{Center for Quantum Information, IIIS, Tsinghua University, Beijing 100084, People's Republic of China}
	%\affiliation{CAS Key Laboratory of Quantum Information,University of Science and Technology of China, Hefei 230026, China}
\begin{abstract}{
{Quantum steering is considered as one of the most well-known nonlocal phenomena in quantum mechanics. Unlike entanglement and Bell non-locality, the asymmetry of quantum steering makes it vital for one-sided device-independent quantum information processing. Although there has been much progress on steering detection for bipartite systems,
the criterion for EPR steering in tripartite systems remains challenging and inadequate.
In this paper, we firstly derive a novel and promising steering criterion for any three-qubit states via correlation matrix. Furthermore, we propose the monogamy relation between the tripartite steering of system and the bipartite steering of subsystems based on the derived criterion. Finally, as illustrations, we demonstrate the performance of the steering criterion and the monogamy relation by means of several representative examples. We believe that the results and methods presented in this work could be
beneficial to capture genuine multipartite steering in the near future.}}
\end{abstract}
\date{\today}
\maketitle

\section{Introduction} %µÚ1½Ú ÒýÑÔ
In 1935, Einstein, Podolsky, and Rosen (EPR) put forward the celebrated paradox in which they pointed out the incompleteness of quantum mechanics, known as EPR paradox \cite{1}. To formalize this argument, Schr\"{o}dinger \cite{2} subsequently introduced the notion of quantum steering. Specifically, it was assumed that Alice and Bob share a maximally entangled state
\begin{align}
\left| {{\psi _{AB}}} \right\rangle = \frac{1}{{\sqrt 2 }}(\left| {01} \right\rangle  - \left| {10} \right\rangle ),
\end{align}
where $\left| {1} \right\rangle$ and $\left| {0} \right\rangle$ denote the two eigenstates of the spin operator ${\sigma _z}$.
Because of the perfect anticorrelations of the above state, if Alice measures her particle with observable ${\sigma _z}$ and obtains the result of $+1$ or $-1$, the state of the corresponding Bob's particle will collapse to $\left| {1} \right\rangle$ or $\left| {0} \right\rangle$.
Whereas, if Alice's measurement choice is the observable ${\sigma _x}$, then the state of Bob's particle will be collapsed to either ${\left| x_+ \right\rangle } = (\left| 0 \right\rangle  + \left| 1 \right\rangle )/\sqrt 2 $ or ${\left| x_- \right\rangle } = (\left| 0 \right\rangle  - \left| 1 \right\rangle )/\sqrt 2$.
Herein, it can be seen that Alice is capable of making another particle instantly collapse into a different state by performing local measurements on her own particle, which is called quantum steering by Schr\"{o}dinger.
In other words, quantum steering is a unique property of quantum systems, which describes the ability to instantaneously influence one subsystem in a two-party system by performing a local measurement of the other.

In order to attempt to interpret incompleteness of quantum mechanics, scientists put forward  a local hidden variable (LHV) model theory \cite{3}. Notably, in 1964, Bell \cite{4} derived the famous Bell inequality by using the LHV model, and found that Bell inequality can be violated in reality.
Note that, the violation of this inequality means that the predictions of quantum theory cannot be explained by the LHV model, revealing the non-locality of quantum mechanics.
At that time the concept of quantum steering had not yet been mathematically defined. It was not until 2007 that Wiseman et al. \cite{5,6} formally introduced the definition of quantum steering. They described quantum steering as a quantum nonlocal phenomenon that cannot be explained by local hidden state (LHS) models.

As a kind of quantum non-locality, quantum steering is different from quantum entanglement \cite{7,8,9} and Bell non-locality. To be explicit, the characteristic of quantum steering is inherent asymmetry \cite{10,11,12,13,14,T7}, and even one-way steering may occur.
In some cases, party $A$ can steer $B$, while $B$ cannot steer $A$ \cite{15,16}.
Therefore, quantum steering, as an effective quantum resource, plays a crucial role in various quantum information processing tasks, such as one-sided device-independent quantum key distribution \cite{17,18,19}, secure quantum teleportation \cite{20,21}, quantum randomness certification \cite{22,23}, subchannel discrimination \cite{T1,T2}, etc.

To judge whether a quantum state is steerable, several authors have significantly contributed to explaining this issue and brought up numerous different criteria \cite{24,25,26,27,28,29,30,31,32,33,34,35,36,37,38,39} of quantum steering.
For example, there have existed the steering criterion based on linear steering inequalities \cite{24,25,26}, the local uncertainty relation \cite{27,28,29,30,31,32}, all-versus-nothing proof \cite{33}, Clauser-Horne-Shimony-Holt-like (CHSH-like) inequalities \cite{34,35,36}, and so on. Then detection of steerability can be achieved by steering robustness \cite{T1}, steerable weight \cite{40}, and violating those various steering inequalities, etc.
In experimental research, several criteria for quantum steering have been verified \cite{25,41,42,43,44,45,46,T6}. A groundbreaking experiment was proposed by Ou et al. \cite{42} in 1992 using Reid¡¯s criterion \cite{39} to demonstrate the existence of quantum steering. To date, many criteria for bipartite steering detection have been proposed. However, there have been few investigations into tripartite steering detection, which still needs to be addressed.

Among those steering criteria of bipartite systems,  Lai and Luo \cite{47}
employed correlation matrix of the local observations and proposed the steerability criterion for bipartite systems of any dimension, and three classes of local measurements, including local orthogonal observables (LOOs) \cite{48}, mutually unbiased measurements (MUMs) \cite{49}, and general symmetric informationally complete measurements (GSICs) \cite{50}, were applied on attaining the proposed steering criteria. Inspired by Lai and Luo's work,   we first derive the steering criterion for an arbitrary three-qubit quantum state via correlation matrices with LOOs. Besides, for a three-qubit state, we also propose a monogamy relation between three-party steering and subsystem two-party steering.

The remainder of the  paper is arranged as follows. Sec. II introduces the notion of EPR steering and several well-known criteria. In Sec. III, we present a new  criterion of EPR steering in tripartite systems and also present its proof. Furthermore, we put forward the monogamy relation between three-party steering and subsystem two-party steering. As illustrations, we render several representative examples to demonstrate the detection ability of our criterion  in Sec. IV. Finally, we conclude the  paper with a summary in Sec. V.
\section{EPR STEERING}
In 2007, Wiseman et al. \cite{5} provided the definition of quantum steering. To be more specific, Alice prepares two entangled particles, sends one  to Bob, and declares that she can steer the state of Bob's particle by measuring her remaining particle. For each measurement choice $x$ and measurement result $a$ of Alice, Bob will gain the corresponding unnormalized conditional state ${\sigma _{a| x }}$. These unnormalized conditional states satisfy $\sum\nolimits_a {{\sigma _{a| x }}}  = {\rho _B}$, which ensures that Bob's reduced state ${\rho _B} = {\mathrm{tr}_A}({\rho _{AB}})$ does not depend on Alice's choice of measurements. Bob then verifies that the unnormalized conditional state can be described as the LHS model
\begin{align}
{\sigma _{a\left| x \right.}} = \int {d\lambda {p_\lambda }{p_C}(a| x,\lambda )} p_\lambda ^B,
\end{align}
where $\lambda$ represents the hidden variable parameter, ${p_C}(a|x,\lambda)$ represents the local response function, and $ p_\lambda ^B$ represents the hidden state. If the conditional state ${\sigma _{a\left| x \right.}}$ can be described by the local hidden state model, then the quantum state ${\rho _{AB}}$ is not steerable; otherwise, it is steerable.

In the experiment, if we use $x$ and $y$ to represent the measurement choices of Alice and Bob, respectively, and use $a$ and $b$ to represent the measurement results obtained by measuring $x$ and $y$ respectively. For a quantum state ${\rho _{AB}}$ that  conforms to the LHS model, the joint probability distribution can be written as
\begin{align}
p(a,b| {x,y}) = \int {d\lambda {p_\lambda }{p_C}(a| x,\lambda )} {p_Q}(b| y, p_\lambda ^B),
\end{align}
where ${p_C}(a| x ,\lambda )$ denotes classical probability, and ${p_Q}(b| {y,} p_\lambda ^B) = \mathrm{tr}({M_{b| y }}p_\lambda ^B)$ denotes quantum probability. If the probability distribution obtained by the experiment cannot obey this formula, then we say that a bipartite state is steerable from Alice to Bob.

As mentioned in the introduction, many steering criteria have been proposed to judge whether a quantum state is steerable. Here we briefly introduce one  typical criterion for arbitrary bipartite systems via correlation matrices  \cite{47}.   Suppose that Alice and Bob share a bipartite state $\rho$ on a Hilbert space, $\mathcal{A} = \{ {A_i}:i = 1,2,...,m\} $ and $\mathcal{B} = \{ {B_i}:i = 1,2,...,n\}$  are the local observables of the two sets of parties $a$ and $b$, respectively. The corresponding correlation matrix can be written as
\begin{align}
C(\mathcal{A},\mathcal{B}\left| \rho  \right.) = ({c_{ij}}),
\end{align}
with
\begin{align}
{c_{ij}} = \rm{tr}(({A_i} \otimes {B_j})(\rho  - {\rho _a} \otimes {\rho _b})).
\end{align}
Lai and Luo proposed and proved that if $\rho$ is unsteerable from Alice to Bob, then
\begin{align}
{\left\| {C(\mathcal{A},\mathcal{B}\left| \rho  \right.)} \right\|_{\mathrm{tr}}} \le \sqrt {{\Lambda _a}{\Lambda _b}},
\end{align}
where
\begin{align}
{\Lambda _a}=& \sum\limits_{i = 1}^m {V({A_i},{\rho _a})}, \nonumber\\
{\Lambda _b}= \mathop {\max }\limits_{{\sigma _b}} (\sum\limits_{j = 1}^n &{{{(\mathrm{tr}{B_j}{\sigma _b})}^2}} )- \sum\limits_{j = 1}^n {{{(\mathrm{tr}{B_j}{\rho _b})}^2}}.
\label{e1}
\end{align}
With respect to the matrix $C$, ${\left\| C \right\|_{\mathrm{tr}}}$ represents the trace norm, i.e., the sum of singular values. Additionally, the conventional variance of $A_i$ in the state $\rho_A$ is given by $V({A_i},{\rho _a}) = \mathrm{tr}(A_i^2{\rho _a}) - \mathrm{tr}{({A_i}{\rho _a})^2}$, and the maximum is over all states $\sigma _b$ on Bob's side.

\section{Detecting EPR Steering for Tripartite Systems via correlation matrices}
Quantum steering describes the ability to instantaneously influence a subsystem in a two-body system by taking a measurement on the other subsystem. For a three-qubit system, if we would like to explore the system's steering, we have to divide the tripartite system into two parties. Here we divide it into two parties $1 \to 2$, then we can consider this system as the ${\mathbb{C}^2} \otimes {\mathbb{C}^4}$ state. Therefore, based on  steering criterion proposed by Lai and Luo, we extend the two-party criterion to three-party version.

A complete set of LOOs $\{ {G_i}:i = 1,2,...,{d^2}\}$ form the orthonormal basis for all operators in the Hilbert space of a $d$-level system, and satisfy the orthogonal relations $\mathrm{tr}({G_i}{G_j}) = {\delta _{ij}}$. For a tripartite system, we divide it into two parties $A$ and $BC$, and in this paper, two sets of LOOs, $G_m^A$ and $G_n^{BC}$, are chosen for $A$ and $BC$ respectively to detect the steerability of $\rho$,
\begin{align}
G_m^A=& \frac{1}{{\sqrt 2 }}\ {\sigma _m}, m\in \{0,1,2,3\},\\
G_n^{BC}= \frac{1}{2}{\sigma _{[\frac{n}{4}]}} &\otimes {\sigma _{(\frac{n}{4})}},  n\in \{0,1,2,...,15\},
\end{align}
where ${\sigma _m}$ are the Pauli matrices, the signs $[\frac{n}{4}]$ and $(\frac{n}{4})$ represent the integer function and remainder function, respectively.

\emph{Theorem 1}. For arbitrary tripartite state $\rho_{abc}  = \frac{1}{8}\sum\limits_{i,j,k = 0}^3 {{\Theta _{ijk}}{\sigma _i} \otimes {\sigma _j} \otimes {\sigma _k}}$, if
\begin{align}
{\left\| \mathbb{M} \right\|_{{\rm{\mathrm{tr}}}}} > \sqrt {(2 - \mathrm{tr}(\rho _a^2))(1 - \mathrm{tr}(\rho _{bc}^2))},
\label{e2}
\end{align}
then $\rho$ is steerable from $A$ to $BC$, where $\mathbb{M}$ is the correlation matrix constructed with two sets of LOOs $G_m^A$ and $G_n^{BC}$,  and ${\Theta _{ijk}} = \mathrm{tr}({\rho}{\sigma _i} \otimes {\sigma _j} \otimes {\sigma _k})$, the matrix elements can be given by
\begin{align}
{M_{mn}} = {\Theta _{m[\frac{n}{4}](\frac{n}{4})}} - {\Theta _{m00}}{\Theta _{0[\frac{n}{4}](\frac{n}{4})}}.
\end{align}

\emph{Proof.} The elements of the correlation matrix $\mathbb{M}$ can be calculated by
\begin{align}
&{M_{mn}}= \mathrm{tr}[({G_m^A} \otimes {G_n^{BC}})(\rho_{abc}  - {\rho _a} \otimes {\rho _{bc}})]
\label{e3}
\end{align}
where the reduced states can be expressed as
\begin{align}
{\rho _a} = {\mathrm{tr}_{bc}}(\rho_{abc} ) = \frac{1}{2}\sum\limits_{i = 0}^3 {{\Theta _{i00}}{\sigma _i}}, \\
{\rho _{bc}} = {\mathrm{tr}_a}(\rho_{abc} ) = \frac{1}{4}\sum\limits_{j,k = 0}^3 {{\Theta _{0jk}}{\sigma _j} \otimes {\sigma _k}}.
\end{align}
With regard to the Pauli matrices, we have $\mathrm{tr}({\sigma _m}{\sigma _n}) = 2{\delta _{mn}}$, and  we substitute these formulas into Eq. (\ref{e3}) and obtain
\begin{align}
&{M_{mn}}= \mathrm{tr}[({G_m^A} \otimes {G_n^{BC}})(\rho_{abc}  - {\rho _a} \otimes {\rho _{bc}})]\nonumber \\
&= \frac{1}{8}\sum\limits_{i,j,k = 0}^3 {({\Theta _{ijk}} - {\Theta _{i00}}{\Theta _{0jk}})} \mathrm{tr}({\sigma _m}{\sigma _i})\mathrm{tr}({\sigma _j}{\sigma _{[\frac{n}{4}]}})\mathrm{tr}({\sigma _k}{\sigma _{(\frac{n}{4})}})\nonumber\\
&={\Theta _{m[\frac{n}{4}](\frac{n}{4})}} - {\Theta _{m00}}{\Theta _{0[\frac{n}{4}](\frac{n}{4})}}. \label{c}
\end{align}
As for the right-hand side of Eq. (\ref{e2}), Ref. \cite{51} has pointed out that a $d$-dimensional single-particle state $\rho '$  meets
\begin{align}
\sum\limits_i {\mathrm{tr}(G_i^2\rho' )}  = d,\label{a}\end{align}
\begin{align}
\sum\limits_i {\mathrm{tr}{{({G_i}\rho' )}^2}}  = \mathrm{tr}({{\rho'} ^2}).\label{b}
\end{align}
Thus, combining Eqs. (\ref{e1}),  (\ref{a}) and (\ref{b}), we get
\begin{align}
{\Lambda _a}= \sum\limits_{i = 1}^m {V({G_i},{\rho _a})}=2 - \mathrm{tr}(\rho _a^2), \label{d}
\end{align}
which belongs to the right-hand side of Eq. (\ref{e2}). One can find out the maximum $\mathop {\max }\limits_{{\sigma _{bc}}} (\sum\limits_{j = 1}^n {{{({\rm{tr}}{G_n^{BC}}{\sigma _{bc}})}^2}} )$ for all possible quantum state $\sigma _{bc}$, namely
\begin{align}
\sum\limits_{j = 1}^n {{{(\mathrm{tr}{G_n^{BC}}{\sigma _{bc}})}^2}}=\mathrm{tr}(\sigma _{bc}^2),
\end{align}
where, ${\sigma _{bc}} = \frac{1}{4}\sum\limits_{i,j = 0}^3 {{{\rm T} _{ij}}{\sigma _i} \otimes {\sigma _j}}$, and ${\rm tr} ({\sigma _{bc}^2})$ is related to the purity of any two-particle states. Incidentally, the maximum of the purity can reach 1.
As a result, we have
\begin{align}
{\Lambda _{bc}}&= \mathop {\max }\limits_{{\sigma _{bc}}} (\sum\limits_{j = 1}^n {{{(\mathrm{tr}{G_n^{BC}}{\sigma _{bc}})}^2}} )- \sum\limits_{j = 1}^n {{{(\mathrm{tr}{G_n^{BC}}{\rho _{bc}})}^2}}\nonumber\\
&=1 - \mathrm{tr}(\rho _{bc}^2). \label{e}
\end{align}
Based on Eqs. (\ref{c}), (\ref{d}) and (\ref{e}),  Eq. (\ref{e2}) has been proofed.

In addition, considering the intrinsic asymmetry of quantum steering,  we
can judge whether $BC$ can steer $A$  relying on the following criterion. First, we can use a commutative operator that can change $\rho_{abc}$ to $\rho_{bca}$. In this case, $\rho_{bca}  = \frac{1}{8}\sum\limits_{i,j,k = 0}^3 {{\Theta _{ijk}}{\sigma _j} \otimes {\sigma _k}}\otimes{\sigma _i} $. If
\begin{align}
{\left\| \mathbb{M}' \right\|_{{\rm{\mathrm{tr}}}}} > \sqrt {(4 - \mathrm{tr}(\rho _{bc}^2))(1 - \mathrm{tr}(\rho _a^2))},
\end{align}
then $\rho$ is steerable from $BC$ to $A$, where $\mathbb{M'}$ is the correlation matrix, and the matrix elements are
\begin{align}
{M'_{nm}} &= \mathrm{tr}[({G_n^{BC}} \otimes{G_m^A})(\rho_{bc a}  - {\rho _{bc}\otimes{\rho _a}})]\nonumber\\
&= {\Theta _{[\frac{n}{4}](\frac{n}{4})m}} -{\Theta _{[\frac{n}{4}](\frac{n}{4})0}}{\Theta _{00 m}}.
\end{align}

Here, \emph{Theorem 1} presents a steering criterion for evaluating the steerability of a tripartite system, and then we define the difference of the left and right sides of the inequality as
\begin{align}
{H_{A \to BC}} = {\left\| \mathbb{M} \right\|_{{\rm{\mathrm{tr}}}}}- \sqrt {(2 - \mathrm{tr}(\rho _a^2))(1 - \mathrm{tr}(\rho _{bc}^2))}.
\label{e11}
\end{align}
Physically, as long as ${H_{A \to BC}}$ is greater than 0, it means that $\rho$ is steerable from $A$ to $BC$. Therefore,  the quantification of the steering can be expressed as
\begin{align}
{S_{A \to BC}} = \max [{H_{A \to BC}},0].
\label{e4}
\end{align}
Canonically, one can get $\rho_{ac}$ or $\rho_{ab}$ when tracing out $B$ or $C$. As a result, the corresponding steering of the states $\rho_{ab}$, $\rho_{ac}$ and $\rho_{bc}$ can be written as
\begin{align}
S_{A \to B} = \max [{H_{A \to B}},0],\nonumber\\
S_{A \to C} = \max [{H_{A \to C}},0],\nonumber\\
S_{B \to C} = \max [{H_{B \to C}},0],
\label{e5}
\end{align}
where,
\begin{align}
{H_{A \to B}} &={\left\| {C(G,G\left| {{\rho _{ab}}} \right.)} \right\|_{\mathrm{tr}}} - \sqrt {{\Lambda _a}{\Lambda _b}},\nonumber\\
{H_{A \to C}} &= {\left\| {C(G,G\left| {{\rho _{ac}}} \right.)} \right\|_{\mathrm{tr}}} - \sqrt {{\Lambda _a}{\Lambda _c}},\nonumber\\
{H_{B \to C}} &= {\left\| {C(G,G\left| {{\rho _{bc}}} \right.)} \right\|_{\mathrm{tr}}} - \sqrt {{\Lambda _b}{\Lambda _c}}.
\label{e22}
\end{align}

\emph{Theorem 2.} Based on the criterion   proposed above (\emph{Theorem 1}),  for any three-qubit pure state,  the monogamy relation can be obtained as
\begin{align}
{S_{A \to BC}} \ge {H_{A \to B}} + {H_{A \to C}}+ {H_{B \to C}}.
\label{e8}
\end{align}

\emph{Proof.} Its proof has been provided in the {\hyperlink{appendixlink}{Appendix}} in details.

\emph{Corollary 1.} By virtue of \emph{Theorem 2}, if $H_{A \to B}$, $H_{A \to C}$ and $H_{B \to C}$ are greater than or equal to 0, the monogamy relation corresponding to the steering of tripartite system can be further generalized as
\begin{align}
{S_{A \to BC}} \ge {S_{A \to B}} + {S_{A \to C}}+ {S_{B \to C}}.
\label{e6}
\end{align}

\emph{Proof.} Logically speaking, if $H_{A \to B}$, $H_{A \to C}$ and $H_{B \to C}$ are all greater than 0, the following relations are held:
\begin{align}
{H_{A \to B}} = {S_{A \to B}};\nonumber\\
{H_{A \to C}} = {S_{A \to C}},\\
{H_{B \to C}} = {S_{B \to C}}.\nonumber
\end{align}
Due to the above equivalence relations, Eq. (\ref{e8}) can be further rewritten as Eq. (\ref{e6}).

\emph{Corollary 2.} On the basis of \emph{Theorem 2}, if $H_{A \to B}$, $H_{A \to C}$ and $H_{B \to C}$ are less than 0, the monogamy relation corresponding to the steering of tripartite system can be further generalized as
\begin{align}
{S_{A \to BC}} \ge {S_{A \to B}} + {S_{A \to C}}+ {S_{B \to C}}.
\label{e7}
\end{align}

\emph{Proof.} If $H_{A \to B}$, $H_{A \to C}$ and $H_{B \to C}$ are less than 0, the steering of $\rho_{ab}$ and $\rho_{ac}$ can be expressed as
\begin{align}
&{S_{A \to B}}= \max [{H_{A \to B}},0]=0,\nonumber\\
&{S_{A \to C}}= \max [{H_{A \to C}},0]=0,\\
&{S_{B \to C}}= \max [{H_{B \to C}},0]=0,\nonumber
\end{align}
and ${S_{A \to BC}} \ge 0$ is held, we thus have that Eq. (\ref{e8}) can be rewritten as Eq. (\ref{e7}). \vskip 0.6cm

%As for the case of one positive and one negative between  ${H_{A \to B}}$ and $ {H_{A \to C}}$, the relationship between ${S_{A \to BC}}$ and ${S_{A \to B}} + {S_{A \to C}}$ cannot be completely determined due to the complexity. From the current image, when the subsystem steering of the random state satisfies the condition of ${{H_{A \to B}} + {H_{A \to C}}}\ge 0$, most of them conform to Eq. (\ref{e6}), but there are also a small part that do not conform to it. What is more special is the case of ${{H_{A \to B}} + {H_{A \to C}}} < 0$. In this case, the steering distribution of the random state is shown in Fig. \ref{fig.2}(c), although the most dots are above the black line, which means ${S_{A \to BC}} > {S_{A \to B}} + {S_{A \to C}}$, but that doesn't hide the fact that there are still some dots below the black line, so we can't quantify them exactly. However, to a certain extent, this also proves that the steering we are exploring is the genuine tripartite steering, which is different from the steering between subsystems but will also be affected by them.

\section{ILLUSTRATIONS}

In what follows, several representative examples will be offered to illustrate the performance of our steering criteria and the monogamy relation, by employing the randomly generated states, generalized Greenberger-Horne-Zeilinger (GHZ) state  and  generalized W state.
\begin{figure}
\centering
{
\includegraphics[height=5.2cm]{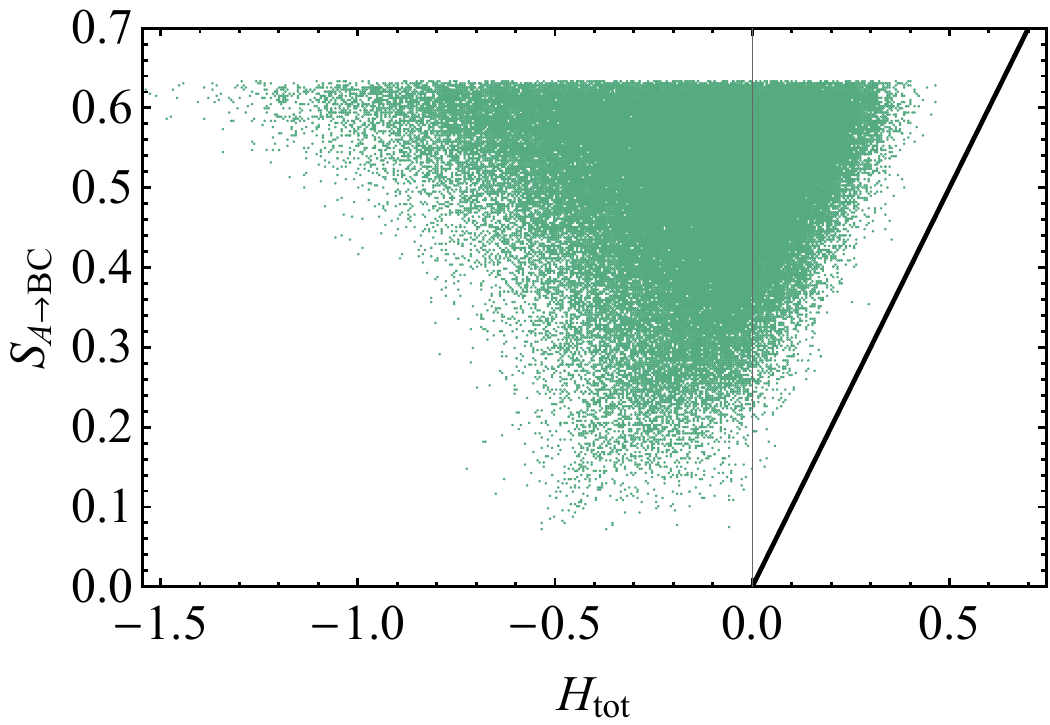}}
\caption{The black line denotes the proportional function with the slope of unity. ${S_{A \to BC}}$ vs ${H_{\rm tot}}$ for ${10^5}$ randomly generated three-qubit pure states, each green dot represents a random pure state.  ${H_{\rm tot}}={H_{A \to B}}+{H_{A \to C}}+H_{B \to C}$ is set.}
\label{fig.1}
\end{figure}

As a matter of fact, there are various effective methods to generate random states \cite{53,54}. Here, we  proceed by introducing the used method for constructing random three-qubit states. It is well established that an arbitrary three-qubit state can be represented by its eigenvalues and normalized eigenvectors as
\begin{align}
\rho  = \sum\limits_{n = 1}^8 {{\lambda _n}\left| {{\Psi _n}} \right\rangle \left\langle {{\Psi _n}} \right|} ; n \in \{ 1,2,3,4,5,6,7,8\},
\end{align}
herein, $\lambda _n$  can be interpreted as the probability that $\rho$ is in the pure state $\left| {{\Psi _n}} \right\rangle$, and the normalized eigenvector of state can establish arbitrary unitary operations $E = \{ {\Psi _1},{\Psi _2},{\Psi _3},{\Psi _4},{\Psi _5},{\Psi _6},{\Psi _7},{\Psi _8}\}$. Thus, an arbitrary three-qubit state can be composed of an arbitrary probability set $\lambda _n$ and an arbitrary $E$. The random number function $\Gamma(a_1, a_2)$ generates a random real number within a closed interval $[a_1, a_2]$. At first, we can generate eight random numbers in this way
\begin{align}
{{\cal N}_1} &= \Gamma(0,1);\ {{\cal N}_2} = {{\cal N}_1}\Gamma(0,1);\nonumber\\
{{\cal N}_3} &= {{\cal N}_2}\Gamma(0,1);\ {{\cal N}_4} = {{\cal N}_3}\Gamma(0,1);\\
{{\cal N}_5} &= {{\cal N}_4}\Gamma(0,1);\ {{\cal N}_6} = {{\cal N}_5}\Gamma(0,1);\nonumber\\
{{\cal N}_7} &= {{\cal N}_5}\Gamma(0,1);\ {{\cal N}_8} = {{\cal N}_7}\Gamma(0,1).\nonumber
\end{align}
The random probability is the set of $\lambda _n \ (n \in \{1,2,3,4,5,6,7,8\})$ controlled by random numbers ${\cal N}_m$, which is expressed as
\begin{align}
{\lambda _n} = \frac{{{{\cal N}_m}}}{{\sum\limits_{m = 1}^8 {{{\cal N}_m}} }}.
\end{align}
In this way, we get a set of random probabilities in descending order. For the random generation of unitary operation, we first randomly give an eight-order real matrix $K$ by the random number function $f(-1, 1)$ with the closed interval $[-1, 1]$. Thus, we can construct a random Hermitian matrix by using the matrix $K$
\begin{align}
H = D + ({U^{\rm T}} + U) + i({L^{\rm T}} - L)
\end{align}
where $D$ denotes the diagonal part of the real matrix $K$, and $L$ ($U$) represents the strictly lower (upper) triangular part of the real matrix $K$, respectively. The superscript $\rm T$ represents the transpose of the corresponding matrix.

By means of this method, one can obtain normalized eigenvectors $\left| {{\Psi _n}} \right\rangle$ of the Hermitian matrix $H$ that forms the random unitary operation $E$. We thereby attain the random three-qubit state $\rho  = \sum\limits_{n = 1}^8 {{\lambda _n}\left| {{\Psi _n}} \right\rangle \left\langle {{\Psi _n}} \right|}$. ${\cal N}_1 = 1$ corresponds to the case of generating three-qubit pure random state.

\emph{Example 1.} By utilizing the above  method, we prepare ${10^5}$ random three-qubit pure states and plot ${S_{A \to BC}}$ versus ${H_{\rm tot}}={H_{A \to B}}+{H_{A \to C}}+H_{B \to C}$  in Fig. \ref{fig.1}. Following the figure,  the green dots corresponding to the ${10^5}$ random states are always above the black slash line with the slope of unity, that is to say, the inequality (\ref{e8}) is held for all the generated random states.

\begin{figure}[htbp]
\centering
{
\includegraphics[width=4.2cm]{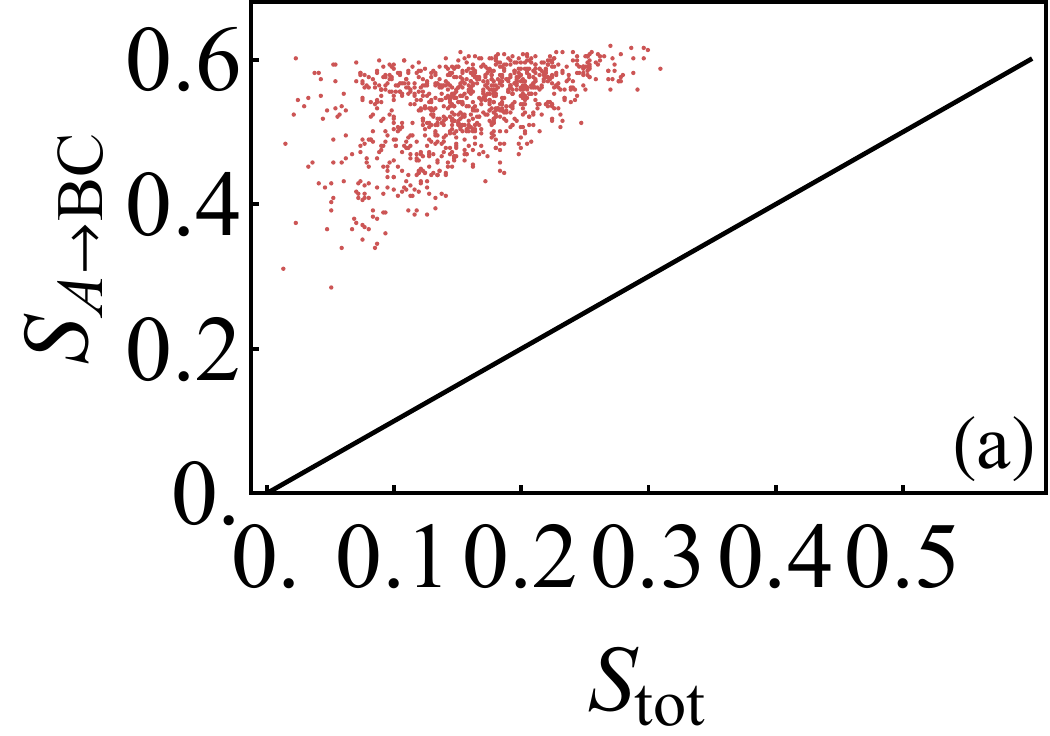}}
{
\includegraphics[width=4.2cm]{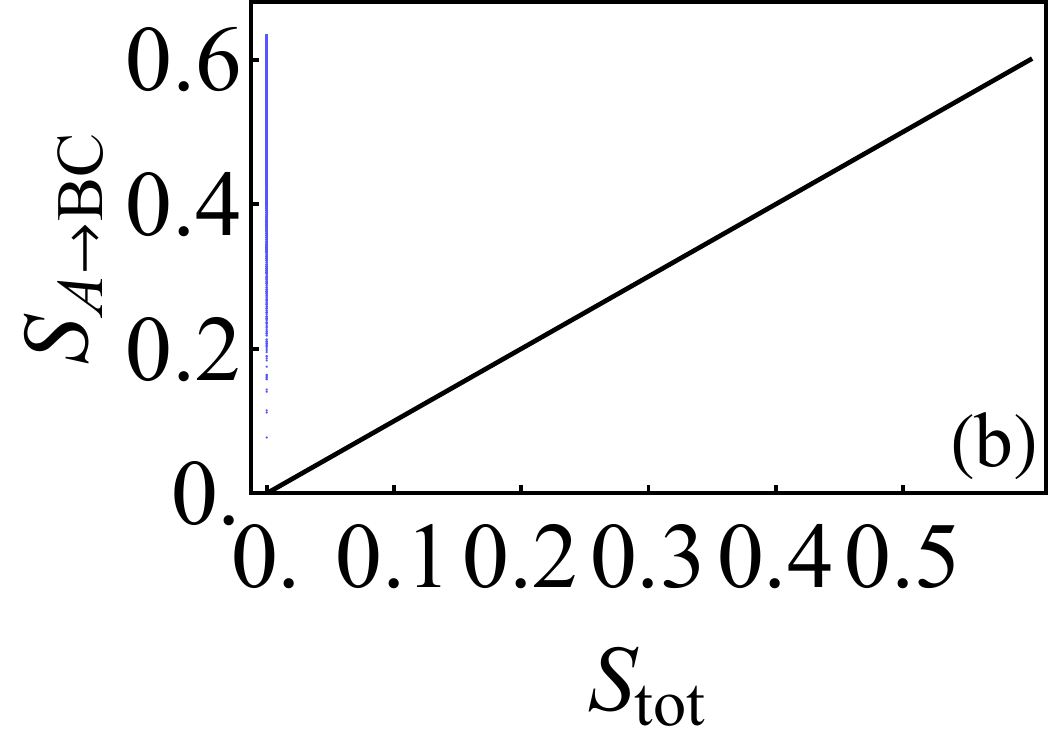}}
\caption{The black line denotes the proportional function with the slope of unity. (a) ${S_{A \to BC}}$ vs $S_{\rm tot}={S_{A \rightarrow B}}+{S_{A \to C}}+{S_{B \to C}}$ for the selected random states, the number of these selected random states, which satisfy $H_{A \to B}\geq 0$, $H_{A \to C}\geq0$ and $H_{B \to C}\geq0$, is 990; (b) ${S_{A \to BC}}$ vs $S_{\rm tot}$ for the selected random states, the number of these selected random states, which satisfy $H_{A \to B}<0$, $H_{A \to C}<0$ and $H_{B \to C}<0$, is 13366. }
\label{fig.2}
\end{figure}
\begin{figure}
\centering
{
\includegraphics[width=8cm]{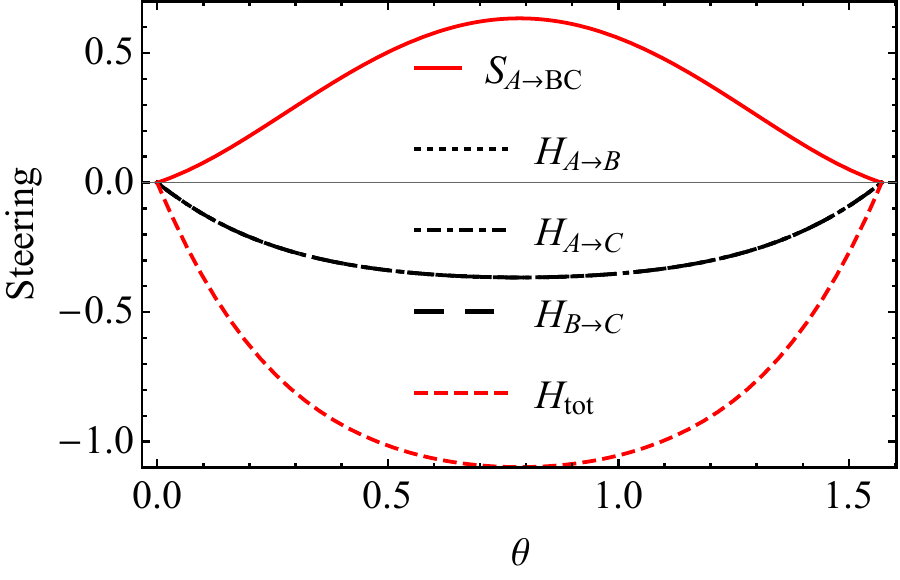}}
\caption{(Color online) Quantum steering vs the state¡¯s parameter $\theta$ in the case of the generalized GHZ state. The black dotted line denotes ${H_{A \to B}}$, the black dot dash line denotes ${H_{A \to B}}$, and he black dashed line denotes $H_{B \to C}$. Specifically, the three black lines are overlapped due to the same expressions (\ref{e10}). the red solid line denotes ${S_{A \to BC}}$, the red dashed line represents $H_{\rm tot}$.}
\label{fig.3}
\end{figure}

\emph{Example 2}. On the basis of \emph{Example 1}, we extract those random states that satisfy the conditions of  \emph{Corollaries 1 or 2}, and draw the steering distribution in Figs. \ref{fig.2}(a) and \ref{fig.2}(b), respectively.
One can easily see that ${S_{A \to BC}}\geq S_{\rm tot}={S_{A \rightarrow B}}+{S_{A \to C}}+{S_{B \to C}}$ is maintained, which essentially displays \emph{Corollaries 1} and \emph{2}.

\emph{Example 3}. Let us first consider a type of three-qubit states, the generalized GHZ states, which can be described  as
\begin{align}
{|\psi\rangle _{\rm GHZ}} = \sin \theta \left| {000} \right\rangle  + \cos \theta \left| {111} \right\rangle,
\end{align}
where $0 \le \theta  \le \pi/2$. Incidentally,  the quantum state will become separable  without any quantum correlation, when $\theta=0\ {\rm or}\ \pi/2$. To determine whether the state is steerable, we can judge by whether it conforms to inequality (\ref{e2}). If the inequality is satisfied, it means that ${\rho _1} = \left| {{\psi _{\rm GHZ}}} \right\rangle \left\langle {{\psi _{\rm GHZ}}} \right|$ is steerable from $A$ to $BC$. According to Eq. (\ref{e2}), we have
\begin{align}
{\left\| {{\mathbb{M}_1}} \right\|_{\rm{tr}}}&=2\left| {\cos \theta \sin \theta } \right|+ 2{\cos ^2}\theta {\sin ^2}\theta, \nonumber\\
2 &- \mathrm{tr}(\rho _a^2)=\frac{1}{4}(5 - \cos4\theta ),\\
1 &- \mathrm{tr}(\rho _{bc}^2) = 2{\cos ^2}\theta {\sin ^2}\theta.\nonumber
\end{align}

For clarity, the variation trend of the corresponding steering ${S_{A \to BC}}$ with the coefficient $\theta$ is plotted in Fig. \ref{fig.3}. As can be seen from Fig. \ref{fig.3}, in the range of $\theta  \in \left[ {0,2\pi } \right]$, ${S_{A \to BC}}$ is always greater than 0, which illustrates the relative tightness of our EPR steering criterion Eq. (\ref{e2}).

On the basis of Eq. (\ref{e5}), the steering of subsystems $\rho_{ab}$, $\rho_{ac}$ and $\rho_{bc}$ can be calculated as
\begin{align}
&{H_{A \to B}}= {H_{A \to C}}= {H_{B \to C}}= 2{\cos ^2}\theta {\sin ^2}\theta  \nonumber \\
-& \sqrt {\left( {2 - \left( {{{\cos }^4}\theta  + {{\sin }^4}\theta } \right)} \right)\left( {1 - \left( {{{\cos }^4}\theta  + {{\sin }^4}\theta } \right)} \right)},
\label{e10}
\end{align}

Fig. \ref{fig.3} has also plotted ${H_{A \to B}}$, ${H_{A \to C}}$, ${H_{B \to C}}$ and ${H_{\rm tot}}$ as a function of the state¡¯s parameter $\theta$. It is interesting to see that $H_{A\rightarrow B}$, $H_{A\rightarrow C}$ and ${H_{B \to C}}$ coincide perfectly, ${H_{A \to B}}, {H_{A \to C}}, {H_{B \to C}}\leq 0$, and ${S_{A \to BC}}\geq {H_{\rm tot}}$ are satisfied all the time, which show the performance of  \emph{Theorem 2} and \emph{Corollary 2} respectively.

\begin{figure}

\centering
{
\includegraphics[width=6.5cm]{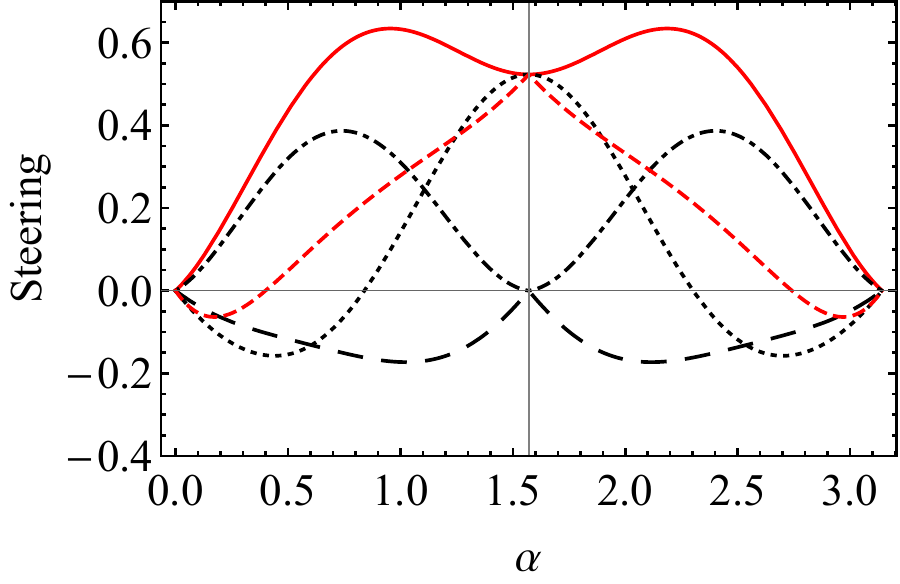}}
{
\includegraphics[width=1.8cm]{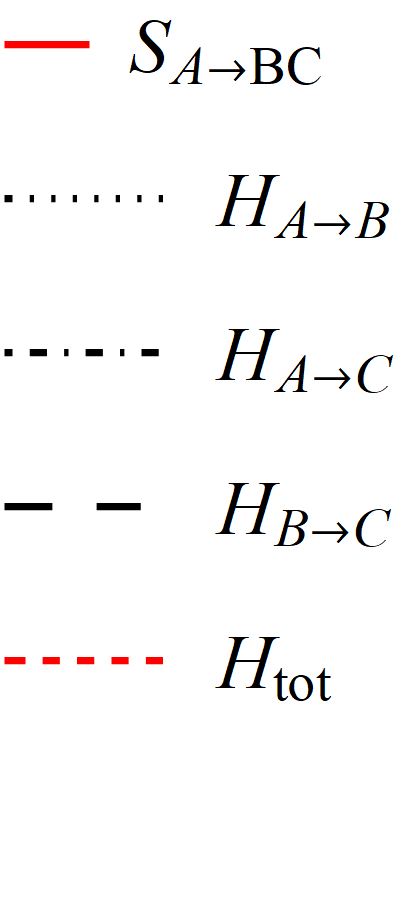}}
\caption{(Color online) Steering of generalized W state vs the state¡¯s parameter $\alpha$, the red solid line represents steering ${S_{A \to BC}}$; the black dotted line denotes ${H_{A \to B}}$, the black dot dash line denotes ${H_{A \to B}}$, and he black dashed line denotes $H_{B \to C}$, the red dashed line represents $H_{tot}$ here which means ${H_{A \to B}}+{H_{A \to C}}+H_{B \to C}$, and the gray vertical line represents $\alpha=\pi/2$.}
\label{fig.4}
\end{figure}

\emph{Example 4}. Let us consider another three-qubit state, the generalized W state, which can be expressed as
\begin{align}
{|\psi\rangle _{\rm W}} = \sin \theta \sin \alpha \left| {100} \right\rangle  + \sin \alpha \cos \theta \left| {010} \right\rangle  + \cos \alpha \left| {001} \right\rangle,
\end{align}
where $\theta  \in \left[ {0,\pi } \right]$ and $\alpha  \in \left[ {0,\pi } \right]$. Without loss of generality, we here choose $\theta  = \frac{\pi }{3}$, hence the two sides of Eq. (\ref{e2}) can be expressed as
\begin{align}
{\left\| {{\mathbb{M}_2}} \right\|_{\rm{tr}}}=&\sqrt {\frac{3}{{8}}(5 + 3\cos2\alpha ){\sin^2}\alpha }\nonumber \\
&+ \left| {\frac{3}{{32}}(5 + 3\cos2\alpha ){\sin^2}\alpha } \right|,\\
2 - \mathrm{tr}(\rho _a^2)=& \frac{1}{{64}}\left( {85 - 12\cos 2\alpha  - 9\cos 4\alpha } \right), \\
2 - \mathrm{tr}(\rho _{bc}^2)=&\frac{3}{{16}}\left( {5 + 3\cos 2\alpha } \right){\sin ^2}\alpha.
\end{align}
Consequently, ${S_{A \to BC}}$ can be drawn as a function of the state's parameter $\alpha$ in Fig. \ref{fig.4}. It is straightforward to see that $S_{A\rightarrow BC}\geq0$, demonstrating the effectiveness of our criterion in detecting the steering for the generalized W state.

In addition, we have the trace norms of the correlation matrix and purities  as
\begin{align}\label{tt1}
&{\left\| {C(G,G\left| {\rho _2^{AB}} \right.)} \right\|_{{\rm{tr}}}} =\frac{{\sqrt 3 }}{2}{\sin ^2}\alpha  + \frac{3}{8}{\sin ^4}\alpha,\nonumber\\
&{\left\| {C(G,G\left| {\rho _2^{AC}} \right.)} \right\|_{{\rm{tr}}}} =\frac{{\sqrt 3 }}{2}\sin2\alpha  + \frac{3}{8}{\sin^2}2\alpha, \nonumber\\
&{\left\| {C(G,G\left| {\rho _2^{BC}} \right.)} \right\|_{{\rm{tr}}}} =\frac{{1 }}{2}\sin2\alpha  + \frac{1}{2}{\sin^2}\alpha{\cos^2}\alpha,\\
&\mathrm{tr}(\rho _a^2)=\frac{1}{{64}}\left( {43 + 12\cos 2\alpha  + 9\cos 4\alpha } \right), \nonumber\\
&\mathrm{tr}(\rho _b^2)=\frac{1}{{64}}\left( {51 + 12\cos 2\alpha  + \cos 4\alpha } \right),\nonumber\\
&\mathrm{tr}(\rho _c^2)={\cos ^4}\alpha  + {\sin ^4}\alpha.\nonumber
\end{align}
By combining Eqs. (\ref{e22}) and (\ref{tt1}), ${H_{A \to B}}$, ${H_{A \to C}}$ and ${H_{B \to C}}$ can be worked out exactly. All the above quantities, ${S_{A \to BC}}$ and ${H_{\rm tot}}$ with respect to the state's parameter $\alpha$ have been depicted in Fig. \ref{fig.4}.
It is apparent that ${S_{A \to BC}}$ (the red solid line) consistently exceeds or equals ${H_{\rm tot}}$ (the red dashed line), and ${H_{A \to B}}$, ${H_{A \to C}}$ and ${H_{B \to C}}$ are greater than or equal to 0, for $\alpha =\pi/2$ (the grey vertical line in the figure). With these in mind, we say that \emph{Theorem 2} and \emph{Corollary 1} are illustrated in the current architecture.
 \vskip 0.6cm

\section{discussion and Conclusion}
Multipartite quantum steering is considered as a promising and significant resource for implementing various quantum communication tasks in quantum networks, which consist of multiple observers sharing multipartite quantum states. In this paper, we have derived the steering criterion for tripartite systems based on the correlation matrix, which might be of fundamental importance in prospective quantum networks. In particular, we utilize LOOs as local measurements to provide operational criteria of quantum steering.

Furthermore, we have put forward the monogamy relation between tripartite steering ${S_{A \to BC}}$ and $H_{\rm tot}$ of the subsystems, based on our derived criterion. It has been proven that ${S_{A \to BC}} \ge {H_{A \to B}} + {H_{A \to C}}+H_{B \to C}$ is always satisfied for arbitrary pure tripartite states.  In addition, we have presented two corollaries in terms of the proposed Theorem.
At the same time, we employed various types of states, including the randomly generated three-qubit pure states, generalized GHZ states and generalized W states, as illustrations for our findings. These examples also show the detection ability of our steering criterion. We believe our criterion provides a valuable methodology for detecting the steerability of any three-qubit quantum states, which may be constructive to generalize into steering criteria for multipartite states in the future.

%As quantum resource, tripartite quantum steering play an important role in secure quantum encryption such as quantum secret sharing (QSS), which is a protocol used to protect highly important messages. The protocol points that Alice sends a message to players (Bob and Charlie), they must collaborate together to obtain the information held by Alice. The effectiveness of protocol is related to the tripartite steering. Our research findings have deepened the understanding of tripartite quantum steering, which may have some effect on QSS.

\section*{Acknowledgements} %ÖÂÐ»£¨ÏîÄ¿Ö§³Ö£©
We would like to express our gratitude to the anonymous reviewer for his/her instructive suggestions. This work was supported by the National Science Foundation of China (Grant Nos. 12075001, 61601002, and 12004006), Anhui  Provincial  Key Research and Development Plan (Grant No. 2022b13020004), Anhui Provincial Natural Science Foundation (Grant No. 1508085QF139), and the fund from CAS Key Laboratory of Quantum Information (Grant No. KQI201701).

\newcommand{\PRL}{\emph{Phys. Rev. Lett.} }
\newcommand{\RMP}{\emph{Rev. Mod. Phys.} }
\newcommand{\PRA}{\emph{Phys. Rev. A} }
\newcommand{\PRB}{\emph{Phys. Rev. B} }
\newcommand{\PRE}{\emph{Phys. Rev. E} }
\newcommand{\PRD}{\emph{Phys. Rev. D} }
\newcommand{\APL}{\emph{Appl. Phys. Lett.} }
\newcommand{\NJP}{\emph{New J. Phys.} }
\newcommand{\JPA}{\emph{J. Phys. A} }
\newcommand{\JPB}{\emph{J. Phys. B} }
\newcommand{\OC}{\emph{Opt. Commun.} }
\newcommand{\PLA}{\emph{Phys. Lett. A} }
\newcommand{\EPJD}{\emph{Eur. Phys. J. D} }
\newcommand{\NP}{\emph{Nat. Phys.} }
\newcommand{\NC}{\emph{Nat. Commun.} }
\newcommand{\EPL}{\emph{Europhys. Lett.} }
\newcommand{\AoP}{\emph{Ann. Phys.} }
\newcommand{\ADP}{\emph{Ann. Phys. (Berlin)} }
\newcommand{\QIC}{\emph{Quantum Inf. Comput.} }
\newcommand{\QIP}{\emph{Quantum Inf. Process.} }
\newcommand{\CPB}{\emph{Chin. Phys. B} }
\newcommand{\IJTP}{\emph{Int. J. Theor. Phys.} }
\newcommand{\IJMPB}{\emph{Int. J. Mod. Phys. B} }
\newcommand{\PR}{\emph{Phys. Rep.} }
\newcommand{\SR}{\emph{Sci. Rep.} }
\newcommand{\LPL}{\emph{Laser Phys. Lett.} }
\newcommand{\OEE}{\emph{Opt. Exp.} }
\newcommand{\IJQI}{\emph{Int. J. Quantum Inf.} }
\newcommand{\EPJC}{\emph{Eur. Phys. J. C} }

\bibliographystyle{plain}

\hypertarget{appendixlink}{}
\appendix
\section*{APPENDIX}

Basically, the Schmidt decomposition for an arbitrary three-particle pure state can be written as form of \cite{55}
\begin{align}
\left| \Psi  \right\rangle  = x\left| {000} \right\rangle  + y{e^{i\phi }}\left| {100} \right\rangle  + z\left| {101} \right\rangle  + h\left| {110} \right\rangle  + \lambda \left| {111} \right\rangle,
\end{align}
where $x^2+y^2+z^2+h^2+\lambda^2=1$ and $x,\ y, \ z,  \ h, \lambda \geq 0$ are maintained. For simplicity, one can set $\phi=0$ and $\lambda=0$.
To prove \emph{Theorem 2}, we first need to prove the following inequality
\begin{align}
{H_{A \to BC}} \ge {H_{A \to B}} + {H_{A \to C}}+ {H_{B \to C}}.
\label{A1}
\end{align}
In accordance to Eqs. (\ref{e11}) and (\ref{e22}), we can obtain ${H_{A \to BC}}$, ${H_{A \to B}}$, ${H_{A \to C}}$ and ${H_{B \to C}}$ for any pure three-qubit state.
If we want to prove that the inequality (\ref{A1}) is valid, we only need to prove that
\begin{align}
{H_{A \to BC}}-({H_{A \to B}} + {H_{A \to C}}+ {H_{B \to C}})\ge 0.
\end{align}
As a result, the left item of the resulting formula can be reexpressed as
\begin{widetext}
\begin{align}\label{A2}
f(x,y,z,h)&=H_{A \to BC}-({H_{A \to B}} + {H_{A \to C}}+ {H_{B \to C}}) \nonumber\\
&= \sqrt 2 z\sqrt {[1 + 2{x^2}({z^2} + {h^2})]({x^2} + {h^2})}  - xh- z(x + h) + \sqrt 2 h\sqrt {[1 + 2{x^2}({z^2} + {h^2})]({x^2} + {z^2})}\\
&+ 2x\sqrt {{z^2} + {h^2}}  + 2{x^2}({z^2} + {h^2}) - zh\sqrt {8zh + {{( - 1 + 2{y^2} + 2zh)}^2}} + \sqrt 2 z\sqrt {[1 + 2{h^2}({z^2} + {x^2})]({x^2} + {h^2})} \nonumber\\
&- \sqrt 2 x\sqrt {[1 + 2{x^2}({z^2} + {h^2})]({z^2} + {h^2})}- \frac{1}{2}(\left| {f + g} \right| + \left| {f - g} \right| + \left| {w + v} \right| + \left| {w - v} \right|)\nonumber
\end{align}
\end{widetext}
with
\begin{widetext}
\begin{align}
f &= x(h - 2{y^2}h + 2{x^2}h), \
g = x\sqrt {{h^2}\left\{1 + 4{y^4} - 4{y^2}(1 + 2xh + {h^2}) - 4h[x + ( - 1 + {z^2})h + {h^3}]\right\}},\nonumber \\
w &= x(z - 2{y^2}z + 2{x^2}z),\
v = x\sqrt {{z^2}\left\{1 + 4{y^4} - 4{y^2}(1 + 2xz + {z^2}) - 4z[x + ( - 1 + {h^2})z + {z^3}]\right\}}.
\end{align}
\end{widetext}
That is to say, if we can prove that $f(x,y,z,h)\geq 0$   in the region of $x^2+y^2+z^2+h^2=1$ with $x \geq 0,\ y \geq 0,\ z \geq 0$ and $\ h \geq 0$, then  the inequality (\ref{A1}) is proven. There are four absolute values in the above formula. In order to facilitate calculation, we can divide the region for removing the absolute value symbols.
As a matter of fact, the region can be  divided  into 16 subregions and the internal and external boundaries.

Besides, we here make use of the Lagrange Multiplier Method to prove inequity (\ref{A1}). If the local minimums of Eq. (\ref{A2}) are greater than 0, it indicates  the desired inequality is true, as the function $f(x,y,z,h)$ is continuous. According to the Lagrange Multiplier Method, to solve  the local minimums of $f(x,y,z,h)$ under condition $x^2+y^2+z^2+h^2-1=0$, we  require constructing the following Lagrange function
\begin{align}
f(x,y,z,h,k)=f(x,y,z,h)-k(-1+x^2+y^2+z^2+h^2),
\end{align}
where $k$ denotes Lagrange multiplier. After then, we take the derivative of $x,\ y,\ z,\ h,\ k$ as
\begin{align}
\left\{ {\begin{array}{*{20}{c}}
{\frac{{\partial f(x,y,z,h,k)}}{{\partial x}} = 0},\\
{\frac{{\partial f(x,y,z,h,k)}}{{\partial y}} = 0},\\
{\frac{{\partial f(x,y,z,h,k)}}{{\partial z}} = 0},\\
{\frac{{\partial f(x,y,z,h,k)}}{{\partial h}} = 0},\\
{\frac{{\partial f(x,y,z,h,k)}}{{\partial k}} = 0},
\end{array}} \right.
\end{align}
respectively. The solutions satisfying these equations are called the critical points. Finally, we choose the critical point satisfying condition $x \geq 0,\ y \geq 0,\ z \geq 0$ and $\ h \geq 0$ to get the local minimum of $f(x,y,z,h)$.

All the subregions and its corresponding critical points' number have been shown in Table~\ref{tab:table1}.  After careful computation, two critical points can be found as
\begin{widetext}\begin{align}
(x,\ y,\ z,\ h) =
\left\{ {\begin{array}{*{20}{l}}
{(0.39036823927218467,\ 0,\ 0.788 617 685 785 144 8,\ 0.47507345056784694)},\\
{(0.39036823927212777,\ 0,\ 0.788 617 685 785 211 7,\ 0.4750734505677587)}.
\end{array}} \right.\nonumber
\end{align}\end{widetext}
Then we can substitute the critical points into Eq. (\ref{A2}), the same local minimum $f_{\rm min}(x,y,z,h)=0.361084$ is obtained, which is obviously greater than 0. In other words, the inequality (\ref{A1}) is held in the subregions.

\begin{table}[b]
\caption{\label{tab:table1}
The specific situation of 16 regions and the number of the critical points of corresponding regions.}
\begin{ruledtabular}
\begin{tabular}{cccccccc}
$f + g$ & $f - g$ & $w+v$ & $w-v$ & The number of critical points \\
\hline
+& + & + & + & 2 \\
+& + & + & - & 0 \\
+& + & - & + & 0 \\
+& - & + & + & 0 \\
-& + & + & + & 0 \\
-& - & + & + & 0 \\
-& + & - & + & 0 \\
+& - & - & + & 0 \\
+& + & - & - & 0 \\
-& + & + & - & 0 \\
+& - & + & - & 0 \\
-& - & + & - & 0 \\
-& + & - & - & 0 \\
+& - & - & - & 0 \\
-& - & - & + & 0 \\
-& - & - & - & 0 \\
\end{tabular}
\end{ruledtabular}
\end{table}

In addition to the critical points within these 16 subregions, there may also exist critical points on the boundaries including internal ones and external ones. Next, let us turn to consider the cases of the region's boundaries.
The internal boundaries refer to those with $f = g$ and  $w = v$, while the external ones refer to those with $x = 0$ or $y = 0$ or $z = 0$ or $h = 0$.
Likewise, we  take advantage of the Lagrange Multiplier Method to judge whether the inequality is valid on the boundaries. In what follows, we will discuss the cases of the internal and external boundaries, respectively.

With respect to  the internal boundary, there exist three cases, i.e.,
\begin{widetext}
\begin{align}
\left\{ {\begin{array}{*{20}{l}}
{\left| {f + g} \right| + \left| {f - g} \right| + \left| {w + v} \right| + \left| {w - v} \right| = 2f + 2w}, \ \  {\rm{for}} \ f = g > 0,\ w = v > 0,\\
{\left| {f + g} \right| + \left| {f - g} \right| + \left| {w + v} \right| + \left| {w - v} \right|{\rm{ }} = 2f - 2w,} \ \ {\rm{for}} \ w = -v < 0,\ f = g > 0,\\
{\left| {f + g} \right| + \left| {f - g} \right| + \left| {w + v} \right| + \left| {w - v} \right| = 2w - 2f,} \ \ {\rm{for}} \ w = v > 0, \ f = -g < 0.
\end{array}} \right.
\end{align}
\end{widetext}
By the Lagrange Multiplier Method, we obtain the critical points shown in Table \ref{tab:table2}.

\begin{table}
\caption{\label{tab:table2}
The number of critical points of boundaries.}
\begin{ruledtabular}
\begin{tabular}{cccccccc}
&$2f+2w$ & $2f-2w$ & $2w-2f$ & $x=0$ & $y=0$  \\
\hline
The number \\
of critical & 0 & 0 & 0 & 2 &0\\
points
\end{tabular}
\end{ruledtabular}
\end{table}

\begin{table}
\caption{\label{tab:table3}
The number of critical points in different cases on the boundary $z=0$.}
\begin{ruledtabular}
\begin{tabular}{cccccccc}
$f_1 + g_1$ & $f_1 - g_1$ & The number of critical points \\
\hline
+& + & 0\\
+& - & 0\\
-& + & 0\\
-& - & 0\\
\end{tabular}
\end{ruledtabular}
\end{table}

\begin{table}
\caption{\label{tab:table4}
The number of critical points in different cases on the boundary $h=0$.}
\begin{ruledtabular}
\begin{tabular}{cccccccc}
$w_1+v_1$ & $w_1-v_1$ & The number of critical points \\
\hline
+& + & 1\\
+& - & 0\\
-& + & 0\\
-& - & 0\\
\end{tabular}
\end{ruledtabular}
\end{table}

For the external boundaries, they can be divided into the following cases. On the boundary with $x=0$, the absolute value items of Eq. (\ref{A2}) will disappear, the function consequently can be simplified into
\begin{align}
 f(y,z,h)&= \sqrt 2 (2zh + zh\sqrt {1 + 2{z^2}{h^2}} )\\
 &\!-\! hz \! \left[1 \!+\! \sqrt {4{y^4} \!+\! {{(1 + 2hz)}^2} \!+ \!{y^2}( \!-\! 4 \!+\! 8hz)} \right]. \nonumber
\end{align}
On the boundary of $y=0$, the corresponding function can be reexpressed as
\begin{widetext}
\begin{align}
f(x,z,h)= &- 4{x^2} + 4{x^4} + \sqrt 2 z \left[ {\sqrt {( { - 1 - 2{h^2} + 2{h^4}} )( { - 1 + {z^2}} )}+ \sqrt {( { - 1 - 2{x^2} + 2{x^4}} )( { - 1 + {z^2}} )} } \right] \nonumber \\
&+ x ( {2\sqrt {1 - {x^2}}  - \sqrt 2 \sqrt {1 + {x^2} - 4{x^4} + 2{x^6}}  -\! 2z}  ) \\
&+ h\left\{ { - 2x + \sqrt 2 \sqrt {( { - 1 + {h^2}})( { - 1 - 2{x^2} + 2{x^4}} )}  - z\left[ {1 + \sqrt {{{( {1 +2hz} )}^2}} } \right]} \right\}.\nonumber
\end{align}
\end{widetext}

To be explicit, we have listed all the critical points of the five cases mentioned above in Table~\ref{tab:table2}. One critical point is
$(x,y,z,h) = (0,\ 0,\ 0.707107,\ 0.707107)$ and we obtain  the local minimum $f_{\rm min}(x,y,z,h)=0.780239$;
and other critical point is $(x,y,z,h) = (0,\ 1,\ 0,\ 0)$ and  $f_{\rm min}(x,y,z,h)=0$ is obtained.
Apparently, all the local minimums are greater than or equal to 0, showing that the inequality (\ref{A1}) is held on these boundaries.

On the boundary of $z=0$ and $h=0$, the function $f$ can be written as
\begin{align}
\!\!f(x,y,h)&\!= \!hx+ 2{h^2}{x^2}\!- \!\frac{1}{2}(\left|f_1 \!+\!g_1\right|\!+\!\left|f_1\! -\!g_1\right|),\\
f(x,y,z)=& (\sqrt 2  + 1)zx + 2{z^2}{x^2} - \frac{1}{2}(\left|w_1 +v_1\right|+\left|w_1 -v_1\right|) ;
\end{align}
with $f_1=x( h + 2{h^2}x - 2h{y^2})$, $w_1=2{x^2}{z^2} + x( z - 2{y^2}z)$, $g_1=x \sqrt {{h^2}[ 1 - 4h( - h + {h^3} + x) - 4( 1 + {h^2} + 2hx){y^2} + 4{y^4} ]}$,  and
$v_1\!=\!x\sqrt {{z^2}[ 1 \!+\! 4{y^4} \!- \!4{y^2}( 1\! +\! 2xz \!+ \!{z^2}) \!- \!4z( x - z + {z^3} )]}$.
The numbers of the critical points on the boundary $z=0$ and $h=0$ have been listed in Tables~\ref{tab:table3} and ~\ref{tab:table4}, respectively. The only critical point  is found as $(x,y,z,h) =(0,\ 1,\ 0,\ 0)$, and we compute that
the corresponding local minimum equals to 0. Thus, the inequality (\ref{A1}) is satisfied on the boundaries with $z=0$ and $h=0$ as well.

In summary, the local minimums of the function $f_{\rm min}(x,y,z,h)\geq 0$ are satisfied all the time in the whole region, consisting of the above 16 subregions and all boundaries. Therefore, the inequality
\begin{align}
{H_{A \to BC}} \ge {H_{A \to B}} + {H_{A \to C}}+ {H_{B \to C}}
\label{ttt1}
\end{align}
is held for all three-qubit pure states.  Owing to Eq. (\ref{e4}), we have $S_{A \to BC}\geq H_{A \to BC}$. Linking with the above inequality (\ref{ttt1}),
the monogamy relation (\ref{e8}) can be obtained as
\begin{align}
{S_{A \to BC}} \ge {H_{A \to B}} + {H_{A \to C}}+ {H_{B \to C}}.
\end{align}
As a consequence, \emph{Theorem 2} has been proved.

\end{document}